# Versa: A Dataflow-Centric Multiprocessor with 36 Systolic ARM Cortex-M4F Cores and a Reconfigurable Crossbar-Memory Hierarchy in 28nm

Sung Kim[1], Morteza Fayazi[1], Alhad Daftardar[1], Kuan-Yu Chen[1], Jielun Tan[1], Subhankar Pal[1], Tutu Ajayi[1],
Yan Xiong[2], Trevor Mudge[1], Chaitali Chakrabarti[2], David Blaauw[1], Ronald Dreslinski[1], Hun-Seok Kim[1]

[1]University of Michigan, Ann Arbor, MI    [2]Arizona State University, Tempe, AZ

**Abstract:** We present Versa, an energy-efficient processor with 36 systolic ARM Cortex-M4F cores and a runtime-reconfigurable memory hierarchy. Versa exploits algorithm-specific characteristics in order to optimize bandwidth, access latency, and data reuse. Measured on a set of kernels with diverse data access, control, and synchronization characteristics, reconfiguration between different Versa modes yields median energy-efficiency improvements of 11.6× and 37.2× over mobile CPU and GPU baselines, respectively.

**Introduction:** Conventional programmable architectures have structurally limited support for diverse, dynamically-varying algorithm characteristics. CPUs provide broad programmability, but are burdened by large out-of-order cores and costly data transfers through fixed-function, coherent caches. GPUs have large SIMD thread arrays and employ scratchpad memories but suffer on irregular workloads. FPGAs are highly configurable but incur high hardware overheads and long reconfiguration times. In contrast, Versa is a highly-parallel architecture that supports regular, irregular, and systolic workloads via fast (2 cycle) reconfiguration. The flexible architecture enables algorithm-specific runtime optimizations. Finally, a tree-based method accelerates multi-threaded barrier synchronization operations. Versa is a standalone flexible architecture, in contrast to coarse-grained systems like [2].

**Architecture Overview:** The Versa architecture (Fig. 1) consists of compute tiles and a 3-level memory hierarchy. Each tile has a cluster of 8 ARM Cortex-M4F worker cores equipped with IEEE 754-compliant floating point units (FPUs). We introduce register-to-register (R2R) links between workers in a 4x2 spatial array, extending transparently across tiles, to support chip-wide systolic algorithms with efficiency gains up to 66.8×. A reconfigurable crossbar (RXB) and 8-slice reconfigurable on-chip memory (ROCM) provide a number of data transfer patterns, each optimized for access latency and throughput. Fast access to cross-mode persistent data (e.g., mutexes) is supported with fixed-function scratchpads in each tile and at the global level. Supervisory tasks and runtime reconfiguration are offloaded to a manager core in each tile.

**Reconfigurable Crossbar and Memory:** The RXB and ROCM (Fig. 2) are co-designed to provide reconfigurable memory slices that are private per worker, shared across all workers in a tile, or FIFO-buffered between core-pairs. The RXB has 1 bidirectional port per worker/ROCM slice and is reconfigurable to either 1) *RXB-shared*, 2) *RXB-private*, or 3) *RXB-queue* modes. RXB-shared provides all-to-all connectivity between workers and ROCM slices with least-recently-granted arbitration [1]. From the perspective of worker cores, shared slices function as single memory that has 8x the capacity of private slices. RXB-shared is also beneficial for workloads where common data is accessed by multiple cores. In RXB-private, RXB crosspoints lock worker-to-slice connections vertically and skip arbitration cycles, reducing access latency by 33%. Privatization eliminates false bank contention with up to ~8× additional improvement in latency and bandwidth. RXB-queue supports common streaming DSP and filtering kernels. Since addresses are unnecessary for streaming FIFO accesses, 'crosspoint splitting' enables simultaneous reads/writes by producer-consumer cores over the same RXB port, and effectively doubled bandwidth.

The L1 ROCM is reconfigurable into cache, SPM, or queue (i.e., FIFO) modes. Cache logic with tags and hit/miss detection is conditionally enabled and data-gated if unused. SRAM banks are fully reused across modes, along with multi-purpose registers that track mode-specific request states (e.g., cache miss handling, FIFO fill levels). 32-bit wide sub-banks match the native bit-width of worker cores. Sub-banking trades 34% area increase for 3.4× lower common-case access energy. RXB-shared and private combine with ROCM cache or SPM, resulting in 2x2 composite configurations in addition to FIFO queue. Mode control is memory-mapped to the manager core, and mode transitions complete in 2 cycles.

**R2R Tunneling:** R2R forms a chip-wide, systolic array configuration where adjacent cores communicate directly, avoiding crossbar and cache overheads (Fig. 3). If enabled at runtime (i.e., in software), the FPU registers s0-s3 are aliased to scalar data links in the <W, E, N, S> directions, respectively. When an instruction writes to an R2R register, data from register writeback - normally directed to the local register file - is instead intercepted by the R2R Shim, and forwarded to an adjacent core. An R2R systolic-write updates the link state, and allows a matching R2R systolic-read to proceed at the neighbor. Systolic-reads from R2R registers proceed normally if the link state is valid. Stall-based flow control prevents stale reads and destructive writes. R2R is tightly-coupled with the M4F core, and flow control is implemented with virtually no overhead by integrating with pipeline stall mechanisms. Similarly, link state (i.e., data valid tracking) requires only 2 bits per link.

**Tree-Based Scratchpad Barriers:** Synchronization barriers are key operations in multi-threaded programs that consume up to 60% of cycles for complex workloads [3]. Versa averts coherence overheads with dedicated scratchpads that provide predictable low-latency access. To reduce the serialized barrier section, scratchpads are placed at the tile (T-SPM) and global (G-SPM) levels, enabling a tree-based strategy (Fig. 4). The centralized scratchpad-based approach alone achieves a 1.7× speedup compared to cache-based barriers from the pthreads library on CPU. The tree-based strategy yields additional 3.8× speedup (6.5× total), despite a 9× higher threadcount.

**Measurement Results:** Versa was tested on MachSuite kernels [4] against a mobile-class 4-core ARM A57 CPU and 256-core Tegra X1 GPU. Stencil2D (2D convolution) and GeMM (matrix mult.) exhibit regular data accesses to dense data, while KMP (string search) and SpMV (sparse matrix-vector mult.) have data-dependent variation in access patterns. Mergesort is a branch and synchronization-heavy comparison-based sort. We selected 2 Versa modes per kernel, and modulated data sizes up to 512 KB.

Across kernels, median energy-efficiency improvements of 11.6× and 37.2× (Fig. 5) are achieved versus the CPU and GPU, respectively. Energy-efficiency improvements between Versa modes extend up to 3.17×, illustrating how reconfiguration captures workload-dependent variation (Fig. 6). For instance, Stencil2D with Versa private cache yields 1.37× higher GFLOPS/W relative to private SPM+R2R at small data sizes, but the advantage between modes is inverted at larger sizes. This result is largely due to the use of R2R to share and reuse overlapped input patches across cores, and cache pressure as dataset footprint increases. On Mergesort, Versa attains 2.33× and 71.6× speedups over the CPU and GPU, respectively, translating to 14.4× and 105× energy-efficiency improvements. GPU profiling indicates bottlenecks in parallel synchronization and branch-heavy comparison operations. Results from Mergesort suggest that Versa's independent scalar cores and tree-based scratchpad barriers are effective in-practice.

At 1.0V nominal voltage, the chip operates at 510 MHz clock frequency resulting in 11.9 GFLOPS peak-practical performance and 810 mW power dissipation (Fig. 7). Voltage scaling to the 0.6V minimum-energy point (MEP) improves energy-efficiency by 2.47× (36.4 GFLOPS/W) while dissipating 7.9 mW at 31 MHz. Energy-per-cycle varies from 543 - 1588 pJ/cycle between 0.6 - 1.0V. Versa was fabricated in 28 nm CMOS (Fig. 8) and occupies 12 mm[2].

**Acknowledgements**: This work was sponsored in part by the U.S. Government under the DARPA SDH program, agreement number FA8650-18-2-7864. We thank ARM for IP and design support.

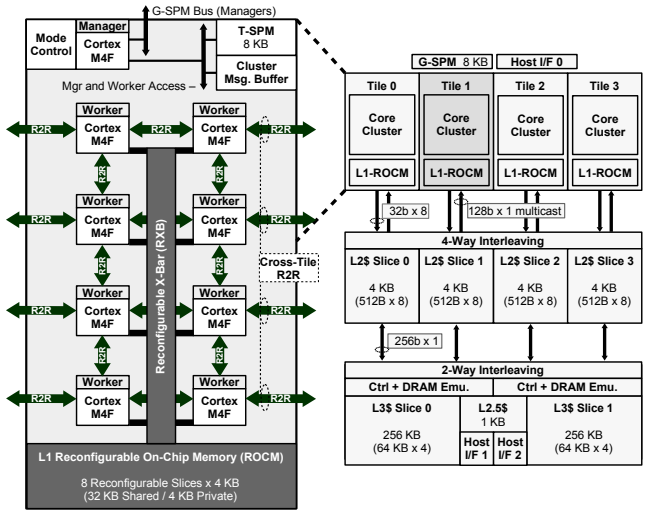

Fig. 1: Main components of a Versa tile (left), and the chip overview (right). Worker cores connected by systolic R2R links and dynamically-reconfigurable memories exploit algorithm-dependent characteristics.

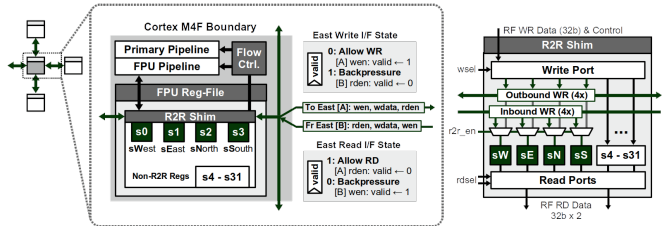

Fig. 3: R2R tunneling and integration in the ARM Cortex M4F.

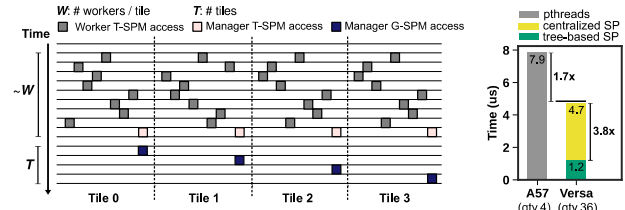

Fig. 4: Illustration of reduced serialization with tree-based barrier synchronization (left), and latency improvements over pthreads (right).

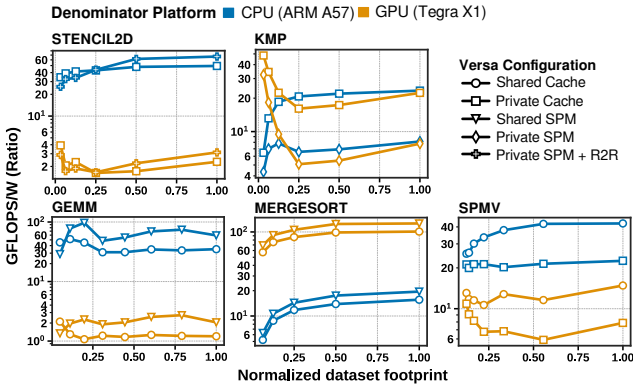

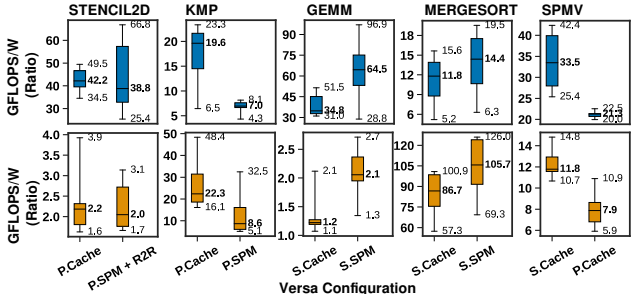

Fig. 6: Trends (top) in energy efficiency across dataset sizes with different Versa configurations, and summary boxplots (bottom).

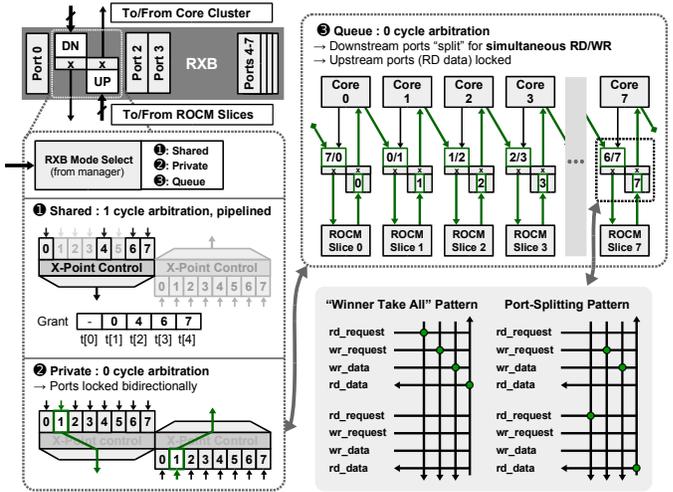

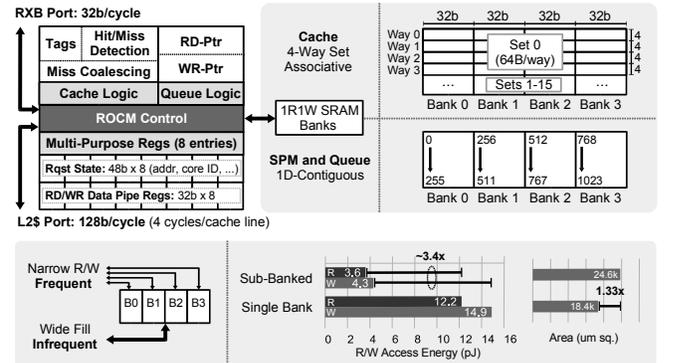

Fig. 2: Shared, private, and queue modes in the reconfigurable crossbar (top), and summary of the L1-ROCM implementation (bottom).

| Kernel | Config A | Config B | Reconfig. Gain | Median / A57 | Median / TX1 |
|---|---|---|---|---|---|
| STENCIL2D | Prvt. Cache | Prvt. SPM+R2R | 1.26x | 6.92x / 42.6x | 0.16x / 2.25x |
| KMP | Prvt. SPM | Prvt. Cache | 2.62x | 3.18x / 19.6x | 1.51x / 22.3x |
| GEMM | Shrd. Cache | Shrd. SPM | 1.73x | 10.5x / 64.5x | 0.15x / 2.18x |
| MERGESORT | Shrd. Cache | Shrd. SPM | 1.22x | 2.33x / 14.4x | 71.6x / 105x |
| SPMV | Prvt. Cache | Shrd. Cache | 1.57x | 5.42x / 33.5x | 1.80x / 11.8x |
| All Kernels | | | 1.53x | 5.86x / 37.2x | 0.78x / 11.6x |

Fig. 5: Median energy-efficiency (GFLOPS/W) and performance (GFLOPS) gained with runtime reconfiguration at nominal voltage. 'Reconfig. Gain' shows the median gain provided by either Config A or Config B, since the optimal mode varies.

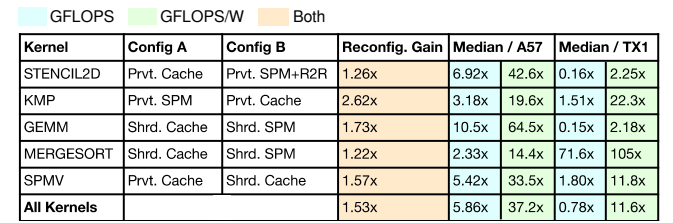

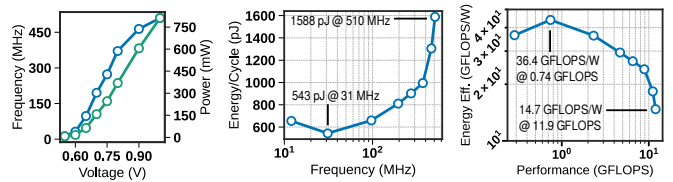

Fig. 7: Power (left), energy-per-cycle (middle), and compute efficiency (right) under VDD scaling between 0.55 - 1.0V. 0.6V coincides with the minimum-energy point.

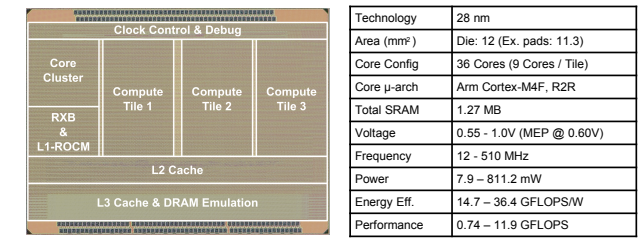

| Technology | 28 nm |
|---|---|
| Area (mm²) | Die: 12 (Ex. pads: 11.3) |
| Core Config | 36 Cores (9 Cores / Tile) |
| Core µ-arch | Arm Cortex-M4F, R2R |
| Total SRAM | 1.27 MB |
| Voltage | 0.55 - 1.0V (MEP @ 0.60V) |
| Frequency | 12 - 510 MHz |
| Power | 7.9 - 811.2 mW |
| Energy Eff. | 14.7 - 36.4 GFLOPS/W |
| Performance | 0.74 - 11.9 GFLOPS |

Fig. 8: Die photo and chip summary.